\documentclass[11pt]{article}
\usepackage{moriond}
\usepackage{amssymb,amsmath,amsbsy,cancel}
\usepackage{tikz}
\usetikzlibrary{arrows,positioning,shapes}
\usetikzlibrary{decorations.pathmorphing}
\usetikzlibrary{decorations.markings}

\tikzset{
fermion/.style={very thick,draw=black, line cap=round, postaction={decorate},
    decoration={markings,mark=at position .65 with {\arrow[black, thick]{triangle 45}}}},
photon/.style={very thick, line cap=round,decorate, draw=black,
    decoration={snake,amplitude=2pt,segment length=8pt}},
boson/.style={very thick, line cap=round,decorate, draw=black,
    decoration={snake,amplitude=2pt,segment length=8pt}},
gluon/.style={very thick,line cap=round, decorate, draw=black,
    decoration={coil,aspect=1,amplitude=3pt, segment length=8pt}},
scalar/.style={dashed, very thick,line cap=round, decorate, draw=black},
ghost/.style={dotted, very thick,line cap=round, decorate, draw=black}
 }

\def\mco{\multicolumn}

\def\vpj{\mbox{${\varphi^\dag i\,\raisebox{2mm}{\boldmath ${}^\leftrightarrow$}\hspace{-4mm} D_\mu\,\varphi}$}}
\def\vpjt{\mbox{${\varphi^\dag i\,\raisebox{2mm}{\boldmath ${}^\leftrightarrow$}\hspace{-4mm} D_\mu^{\,I}\,\varphi}$}}

\def\lcal{{\cal L}}

\def\be{\begin{equation}}
\def\ee{\end{equation}}
\def\bea{\begin{eqnarray}}
\def\eea{\end{eqnarray}}

\newcommand{\Photo}{}

\begin{document}
\vspace*{4cm}
\title{ Lepton Flavor Violation in the Standard Model with dimension 6 operators }

\author{Saereh Najjari \footnote{e-mail: saereh.najjari@fuw.edu.pl}}

\address{Institute of Theoretical Physics, University of Warsaw,
  00-681 Warsaw, Poland}

\maketitle\abstracts{We investigate Lepton Flavor Violation (LFV) in
  the general extension of the Standard Model parametrized by the
  gauge invariant operators of dimension-6. We discuss 3-body charged
  lepton decays and decay of $Z$ boson to lepton pair and compare the
  obtained analytical expression with the current experimental
  results. We derive the numerical bounds on the size of the Wilson
  coefficients parameterizing the allowed size of New Physics
  effects.}

\section{Parametrization of the SM extensions in terms of effective operators}

The renormalizable Standard Model (SM) is probably an effective theory
valid only up to some energy scale where a more fundamental theory
could manifest itself. At the electroweak scale the effects of New
Physics (NP) can be effectively parametrized by new interactions described
by non-renormalizable operators of higher mass dimensions.
Coefficients of such operators (called Wilson coefficients) are
suppressed by heavy mass scale $(\Lambda)$ at which New Physics should become
effective. In general the dimension-4 renormalizable SM Lagrangian can
be extended as follows:
\bea
\label{Leff}
\lcal_{\mathrm {SM}} = \lcal_{\mathrm {SM}}^{(4)} + \frac{1}{\Lambda }
\sum_{k} C_k^{(5)} Q_k^{(5)} + \frac{1}{\Lambda^2} \sum_{k} C_k^{(6)}
Q_k^{(6)} + \ldots\,.
\eea
where $Q_k^{(n)}$ denote the higher-dimension operators and
$C_k^{(n)}$ stand for the corresponding dimensionless Wilson
coefficients.

Using such Lagrangian, the theoretical calculations of relevant
physical observables can be performed in a model-independent way, with
final formulae given in terms of Wilson coefficients.  Having such
expressions simplifies significantly a comparison of various SM
extension with the experimental results, as now only the values Wilson
coefficients of new operators need be calculated within a given model
of NP - this part of analysis is always model-dependent.

In the extended SM with the neutrino mass term, the GIM mechanism makes the branching ratio of the charged lepton flavor violating (CLFV) very small due to smallness of the mass of the neutrino comparing to the mass of the heaviest particle in the loop. Experimentally, the CLFV decays have never been observed yet but there are many models beyond the SM predict sizable rates up to the current experimental bounds. Here, we investigate the $\ell_i \to \ell_j\ell_j\ell_j$ decay and $Z\to \ell_f^+\ell_i^- $ decays in the extension of the SM with dimension 6 operators.The full list of operators of dimension 5 and 6 which can be
constructed out of SM fields is given in \cite{Grzadkowski:2010es}.

 After simplification only 8 operators of dimension 6 give contribution to these decays at tree level:
\begin{itemize}
\item 2 dipole-type operators, contributing to tree-level flavor
  non-diagonal $Z$ and photon couplings to leptons:
\\ \\
$Q_{eW} = (\bar \ell_i\sigma^{\mu\nu}e_j)\tau^I \varphi
  W_{\mu\nu}^I$,~~ $Q_{eB}= (\bar \ell_i \sigma^{\mu\nu} e_j) \varphi
  B_{\mu\nu}$.
\\
\item 3 operators modifying tree-level $Z$ and $W$ couplings:
\\ \\
$Q_{\varphi\ell}^{(1)}=(\vpj)(\bar\ell_i\gamma^\mu
  \ell_j)$,~~~$Q_{\varphi\ell}^{(3)}= (\vpjt)(\bar \ell_i \tau^I
  \gamma^\mu\ell_j)$,~~$Q_{\varphi e}=(\vpj)(\bar e_i \gamma^\mu
  e_j)$
\\
\item 3 four-lepton contact operators:
\\ \\
$Q_{\ell e}=(\bar \ell_i \gamma_\mu \ell_j)(\bar e_k \gamma^\mu
  e_l)$,~~$Q_{\ell \ell }=(\bar {\ell_i} \gamma_\mu
  \ell_j)(\bar\ell_k\gamma^\mu \ell_l)$,~~ $Q_{ee}=(\bar e_i
  \gamma_\mu e_j)(\bar e_k \gamma^\mu e_l)$
\end{itemize}

In~\cite{Crivellin:2013hpa} we list the Feynman rules arising from dimension 6 operators.
\section{Lepton Flavor violation in 3-body charged lepton and $Z$ to lepton pair decays}
\label{sec:numal}

 In this section we compare the analytical results with experimental bounds on the $\ell_i \to \ell_j\ell_j\ell_j$ decay and $Z\to \ell_f^+\ell_i^- $ decays to constrain the Wilson coefficients.Such decays can be generated already at tree level. Our results for radiative lepton decays are given
in~\cite{Crivellin:2013hpa,Najjari:2013una}.   The
experimental bound on these decays are given in
Tables~\ref{table:3lexp} and~\ref{table:Zll}.

\begin{table}[htb]
\centering
\begin{tabular}{|c|c|}
  \hline\hline
\mco{1}{|c|}{Process}   &\mco{1}{|c|}{Experimental bound}
\\ \hline
${\cal B}\left[\tau^-\to\mu^-\mu^+\mu^-\right] $ & ~$2.1\times
10^{-8}$ ~\cite{Hayasaka:2010np} \\
\hline
${\cal B}\left[\tau^-\to e^-e^+e^-\right] $ & ~$2.7\times
10^{-8}$ ~\cite{Hayasaka:2010np} \\

\hline
$ {\cal B}\left[\mu^-\to e^-e^+e^- \right]$ & ~$ 1.0\times
10^{-12}$~\cite{Bellgardt:1987du} \\
\hline \hline
\end{tabular}
\caption{Experimental upper limits on the branching ratios of the
  three body charged lepton decays.\label{table:3lexp}}
\end{table}

\begin{table}[htb]
\centering \vspace{0.8cm}
\renewcommand{\arraystretch}{1.2}
  \begin{tabular}{@{}|c|c|}
\hline
Process & Experimental bound\\
\hline \hline
$\mathrm{Br}\left[Z^0\to\mu^\pm e^\mp\right] $ & ~$1.7 \times 10^{-6}$
~\cite{Abreu:1996mj} \\
\hline
$\mathrm{Br}\left[Z^0\to\tau^\pm e^\mp\right] $ & ~$9.8 \times
10^{-6}$ ~\cite{Abreu:1996mj} \\
\hline
$\mathrm{Br}\left[Z^0\to\tau^\pm \mu^\mp\right] $ & ~$1.2 \times
10^{-5}$ ~\cite{Abreu:1996mj} \\
\hline \hline
\end{tabular}
\caption{Experimental upper limits (95 \% CL) on the lepton flavor
  violating $Z$ decay rates.}
\label{table:Zll}
\end{table}

In general, at the tree-level diagrams mediated by photon, $Z^0$ boson
and 4-lepton contact interactions can contribute to the three body
charged lepton decays $\mu\to 3e$, $\tau\to 3e$ and $\tau\to 3\mu$.
The general expression for the Br($\ell_i\to 3\ell_j$) reads as:
\bea
\mathrm{Br}(\ell_i\to \ell_j \ell_j \bar \ell_j) &=& \frac{m_{\ell_i}^5}{12288
  \pi^3\Lambda^4 \Gamma_{\ell_i}} \left(4 \left(|C_{VLL}|^2 +
|C_{VRR}|^2 + |C_{VLR}|^2 + |C_{VRL}|^2\right)\right.\nonumber\\
&+& |C_{SLR}|^2+ |C_{SRL}|^2 + 48  X_\gamma
\label{eq:br4l}
\eea
where $\Gamma^{\ell_i}$ is the total decay width of the initial
lepton. The $X_\gamma$ (photon contribution) and $C_X$ ($Z$ and
contact interactions) can be written in terms of Wilson coefficients
of operators listed in previous Section as:
\bea
X_\gamma & =& -\frac{16ev}{m_{\ell_i}}\mathrm{Re}\left[\left(2 C_{VLL} +
  C_{VLR} - \frac{1}{2} C_{SLR} \right) C_{\gamma R}^\star + \left( 2
  C_{VRR} + C_{VRL} - \frac{1}{2} C_{SRL} \right) C_{\gamma L}^\star
  \right] \nonumber\\
&+& \frac{64 e^2v^2}{m_{\ell_i}^2}
\left(\log\frac{m_{\ell_i}^2}{m_{\ell_j}^2} - \frac{11}{4}
\right)(|C_{\gamma L}|^2 + |C_{\gamma R}|^2)
\eea
\bea
C_{VLL} &=& 2\left( (2s_W^2-1) \left( C_{\varphi \ell}^{(1)ji} + C_{\varphi
  \ell}^{(3)ji} \right) + C_{\ell\ell}^{jijj} \right) \nonumber\\
C_{VRR} &=& 2 \left( 2 s_W^2 C_{\varphi e}^{ji} +
C_{ee}^{jijj} \right)\nonumber\\
C_{VLR} &=& -\frac{1}{2} C_{SRL} = \left( 2s_W^2 \left( C_{\varphi
  \ell}^{(1)ji} + C_{\varphi \ell}^{(3)ji} \right) + C_{le}^{jijj}\right)
\nonumber\\
C_{VRL} &=& -\frac{1}{2} C_{SLR} = \left( (2s_W^2-1) C_{\varphi e}^{ji} +
C_{\ell e}^{jjji}\right) \nonumber\\
C_{\gamma L}^{ij}&=& C_{\gamma R}^{ji\star} = 2\sqrt{2}s_W \left(c_W
C_{eB}^{ij\star} - s_W C_{eW}^{ij\star} \right)
\eea

 Knowing that $C_{\gamma}$  must be negligible, we  neglect contribution from
$C_{\gamma}$ in order to constrain the Wilson coefficients by using the bounds on $\ell_i \to \ell_j\ell_j\ell_j$ decays.  Normalizing their
branching ratio to the limits in Table~\ref{table:3lexp} we find
following numerical equations constraining the Wilson coefficients:
\bea
{C_{\mu eee}} &\le& 3.29 \times {10^{ - 5}}{\left( {\frac{\Lambda }{{1\;{\rm{TeV}}}}} \right)^2}\sqrt {\frac{{{\rm{Br}}\left[ {\mu  \to eee} \right]}}{{1 \times {{10}^{ - 12}}}}} \,,\nonumber\\
{C_{\tau eee}} &\le& 1.28 \times {10^{ - 2}}{\left( {\frac{\Lambda }{{1\;{\rm{TeV}}}}} \right)^2}\sqrt {\frac{{{\rm{Br}}\left[ {\tau  \to eee} \right]}}{{2.7 \times {{10}^{ - 8}}}}} \,,\\
{C_{\tau \mu \mu \mu }} &\le& 1.13 \times {10^{ - 2}}{\left( {\frac{\Lambda }{{1\;{\rm{TeV}}}}} \right)^2}\sqrt {\frac{{{\rm{Br}}\left[ {\tau  \to \mu \mu \mu } \right]}}{{2.1 \times {{10}^{ - 8}}}}} \,,\nonumber
\label{llll_constraints}
\eea
with ${C_{{\ell _i}{\ell _j}{\ell _j}{\ell _j}}}$ given by
\bea
\begin{array}{l}
{C_{{\ell _i}{\ell _j}{\ell _j}{\ell _j}}} =
 \left( {    2{{\left|     {{\rm{C}}_{\ell e}^{jijj} - 0.54\left( {{\rm{C}}_{\varphi \ell
          }^{\left( 1 \right)ji}    + {\rm{C}}_{\varphi \ell }^{\left( 3
            \right)ji}}   \right)}      \right|}^2}}

 + 2{{\left| {{\rm{C}}_{ee}^{jijj} +  0.46{\rm{C}}_{\varphi e}^{ji}} \right|}^2}

\right.\\ \left. {\;\;\;\;\;\;\;\;\;\;\;\;\;\;\;\;\;

 +{{\left|
      {{\rm{C}}_{\ell e}^{jijj}}+0.46\left( {{\rm{C}}_{\varphi \ell }^   {\left( 1 \right)ji}    +
          {\rm{C}}_{\varphi \ell }^{\left( 3 \right)fi}} \right)
             \right|}^2}

+ {{\left|{{\rm{C}}_{\ell e}^{jjji} - 0.54{\rm{C}}_{\varphi e}^{ji}}
      \right|}^2}

           } \right)^{\frac{1}{2}}.
\end{array}
\label{C_llll}
\eea

Similar analysis can be done for the LFV $Z\to \ell_f^+\ell_i^- $
decays. At the tree level the branching ratio is:
\bea
\mathrm{Br}\left( {Z^0 \to \ell_f^\pm \ell_i^\mp } \right) = \frac{m_Z}{24\pi \Gamma_Z}
\left[\frac{m_Z^2}{2}\left( \left| C_{fi}^{ZR} \right|^2 + \left|
  C_{fi}^{ZL} \right|^2 \right) + \left| \Gamma _{fi}^{ZL} \right|^2 +
  \left| \Gamma _{fi}^{ZR} \right|^2 \right]\,,
\label{eq:Zll}
\eea
with $\Gamma_Z\approx 2.495$~GeV being the total decay width of $Z$
boson.

where we included all tree-level contributions and
\bea
\Gamma^{ZL}_{fi} &=&\dfrac{e}{2 s_W c_W}\left( \dfrac{v^2}{\Lambda^2}
\left( C_{\varphi l}^{(1)fi} + C_{\varphi l}^{(3)fi} \right) + \left( 1 -
2s_W^2 \right) \delta_{fi} \right)\,,\\
\Gamma^{ZR}_{fi} &=& \dfrac{e}{2s_W c_W}\left( \dfrac{v^2}{\Lambda^2}
C_{\varphi e}^{fi} - 2s_W^2 \delta_{fi} \right)\,,\\
C^{ZR}_{fi} &=& C^{ZL\star}_{if} =  -v\sqrt{2} C_Z^{fi}
\eea
where $C_Z^{fi}$ is defined as
\bea
C_Z^{fi} = \left(s_W C_{eB}^{fi} + c_W C_{eW}^{fi} \right) \,.
\label{eq:cz}
\eea

In the experimental values for branching ratio are for the sum $Z\to \ell_f^\pm  \ell_i^\mp+\ell_f^\mp \ell_i^\pm$ ~\cite{Illana:1999ww} while in this equation branching ratio is for the decay $Z\to \ell_f^\pm  \ell_i^\mp$ or $\ell_f^\mp \ell_i^\pm$. Therefore this equation must be multiply by a factor 2 in order to compare into experimental value.

Again normalizing the formulae for the branching ratios to the current
experimental bound listed in Table~\ref{table:Zll} we derived the
numerical equations constraining the Wilson coefficients contributing
to this decay:
\begin{align}
\sqrt{\left| C_{\varphi \ell}^{(1)12} + C_{\varphi \ell}^{(3)12} \right|^2 +
  \left| C_{\varphi e}^{12} \right|^2+ \left| C_Z^{12} \right|^2+ \left|
  C_Z^{21} \right|^2 } &\le 0.06 \left( \frac{\Lambda}{1\;\rm{TeV}}
\right)^2 \sqrt{\dfrac{{\rm Br}\left[ Z^0 \to \mu^\pm e^\mp
      \right]}{1.7\times 10^{-6}}}\,,\nonumber\\{}
\sqrt{\left| C_{\varphi \ell}^{(1)13} + C_{\varphi e}^{13} \right|^2 +
  \left| C_{\varphi e}^{13} \right|^2+ \left| C_Z^{13} \right|^2+ \left|
  C_Z^{31} \right|^2 } &\le 0.14 \left( \frac{\Lambda}{1\;\rm{TeV}}
\right)^2 \sqrt{\dfrac{{\rm Br}\left[ Z^0 \to \tau^\pm e^\mp
      \right]}{9.8\times 10^{-6}}}\,
\label{eq:zll_limits} \,,\nonumber\\{}
\sqrt{\left| C_{\varphi \ell}^{(1)23} + C_{\varphi \ell}^{(3)23} \right|^2 +
  \left| C_{\varphi e}^{23} \right|^2+ \left| C_Z^{23} \right|^2+\left|
  C_Z^{32} \right|^2 } &\le 0.16 \left( \frac{\Lambda}{1\;\rm{TeV}}
\right)^2 \sqrt{\dfrac{{\rm Br}\left[ Z^0 \to \tau^\pm \mu^\mp
      \right]}{1.2\times 10^{-5}}}\,.\nonumber\\{}
\end{align}

These constraints are less stringent than the ones from $\ell_i \to \ell_j\ell_j\ell_j$ decays.

\section{Conclusions}

We have calculated the rates of several lepton flavor violating decays
in the Standard Model extended with dimension 6 operators. We present
numerical expressions constraining the relevant Wilson coefficients,
based on experimental bounds for the $\ell_i \to \ell_j\ell_j\ell_j$ decay and $Z\to \ell_f^+\ell_i^- $ decays. There are also some experiments searching for charged lepton flavor violation and are going to upgrade the sensitivity. Observation of charged lepton flavor violation at experiment would be a clear hint for physics beyond the standard model. We show that the bounds on lepton flavor violation couplings are already very strong if the scale of New Physics is low.

\section*{Acknowledgments}

This work has been done in collaboration with A. Crivellin and J. Rosiek. I acknowledges financial support from the organizing
committee to attend the conference.

\end{document}